\documentclass[aps,prb,twocolumn,showpacs,superscriptaddress]{revtex4-1}

\usepackage{amsmath,amssymb}
\usepackage{bm}
\usepackage{verbatim}
\usepackage{graphicx}
\usepackage{dcolumn}
\usepackage{amstext}
\usepackage{mathptmx}
\usepackage{verbatim}
\usepackage[colorlinks,citecolor=blue,bookmarks]{hyperref}
\usepackage{times}

\begin{document}

\title{Fermi Surface and Band Structure of BiPd from ARPES Studies}
\author{H. Lohani}
\affiliation{Institute of Physics, Sachivalaya Marg, Bhubaneswar- 751005, India.}
\author{P. Mishra}
\altaffiliation[Current address:]{Tata Institute of Fundamental Research,\\
Homi Bhabha Road, Mumbai 400005, India }
\affiliation{Institute of Physics, Sachivalaya Marg, Bhubaneswar- 751005, India.}
\author{B. R. Sekhar}
\email[]{sekhar@iopb.res.in}
\affiliation{Institute of Physics, Sachivalaya Marg, Bhubaneswar- 751005, India.}

\author{Anurag Gupta}
\affiliation{National Physical Laboratory(CSIR),\\
Dr. K. S. Krishnan Road, New Delhi 110012, India.}
\author{V. P. S. Awana}
\affiliation{National Physical Laboratory(CSIR),\\
Dr. K. S. Krishnan Road, New Delhi 110012, India.}

\date{\today}
\begin{abstract}

We present a detailed electronic structure study of the non-centrosymmetric 
superconductor BiPd based on our angle resolved photoemission spectroscopy 
(ARPES) measurements and Density Functional Theory (DFT) based calculations. 
We observe a high intensity distribution on the Fermi surface (FS) of this 
compound resulting from various electron and hole like bands which are present 
in the vicinity of the Fermi energy (E$_f$). The near E$_f$ states are 
primarily composed of Bi-6p with a little admixture of Pd-4d$_{x^2-y^2/zy}$ 
orbitals. There are various spin-orbit split bands involved in the crossing of 
E$_f$ making a complex FS. The FS mainly consists of multi sheets of three 
dimensions which disfavor the nesting between different sheets of the FS.  Our 
comprehensive study elucidates that BiPd is essentially a s-wave multiband 
superconductor.
\end{abstract}

\maketitle

\section{Introduction}

An upsurge has been witnessed recently in the search for novel materials after 
realizing the significant role of spin-orbit coupling (SOC) effects in the 
modification of near Fermi level (E$_f$) electronic structure of materials and 
thereby their physical properties. For example, presence of a strong SOC 
produces conducting edge states in topological insulators 
(TIs)\cite{Hasen1,Zhang}. Similarly, intertwining of the spin-orbit 
interaction with non-centrosymmetric (NCS) structures gives rise to some exotic 
phenomena of mixing up of spin-singlet and triplet Cooper pairing 
channels\cite{Gor,Sig} in superconductors (SCs). The anomalous value of upper 
critical field (H$_{c2}$)\cite{Jha,Zhang1}, presence of Majorana surface 
states at the junction of superconducting transition temperature 
(T$_c$)\cite{maj,maj1} and existence of Weyl fermion surface states in Wyle 
semimetals\cite{weyl} are a few more interesting properties related to the NCS 
structures behaving under SOC effects. These new class of materials not only 
present intriguing physics but also have tremendous scope in various 
applications.

One of the interesting aspects of the NCS crystals is a broken inversion 
symmetry that gives rise to antisymmetric spin-orbit interaction (ASOC) which 
has been theoretically predicted to form an unconventional pairing in the NCS 
SCs. However, most of the NCS SCs, like Mg$_{10}$Ir$_{19}$B$_{16}$\cite{Cava}, 
Mo$_3$Al$_2$C\cite{Kim}, Re$_{24}$Nb$_5$\cite{Lue}, Re$_3$W\cite{Yan} show 
conventional s-type superconducting behavior which is attributed to the weak 
SOC in these compounds. The paring mechanism becomes quite complex due to 
strong electron correlation effects in some other SCs of the NCS family, like 
CePt$_3$Si\cite{Oda}, UIr\cite{Aka}. In this scenario, discovery of 
superconductivity in NCS compound BiPd has brought some new excitation to this 
field due to the presence of heavy elements Bi (Z = 83) and Pd (Z = 46). SOC 
is expected to be strong while electronic correlation is moderate in this 
compound as suggested by Kadowaki-Woods value estimated from resistivity 
measurements\cite{awana}. Therefore, it provides an excellent ground to study 
the role of SOC effects in the electronic structure of NCS SCs. BiPd shows a 
transition from orthorhombic ($\beta$-BiPd) to monoclinic ($\alpha$-BiPd) 
structure at 210$^{\circ}$C and its T$_c$ is $\sim$ 3.7 K\cite{Bhanu,Bhatt}. 
Various measurements, like electrical resitivity, magnetic susceptibility and 
heat capacity helped in establishing a s-wave type BCS superconductivity in 
this system\cite{Bhanu,Sun}. However, mixing of spin-singlet and triplet order 
parameter, signature of multigap and Andreev bound states have also been 
identified from Andreev spectroscopic measurements\cite{Mintu}. Spin-triplet 
component has also been seen in the nuclear quaderpole resonance (NQR) 
measurements\cite{Bhanu1}. Similarly, London penetration depth and its 
corresponding superfluid density showed an anisotropy with two energy gaps 
which was further attributed to the mixing of the two pairing states in 
BiPd\cite{Jiao}. But, so far, there is no consensus reached on the role of 
ASOC induced spin-triplet pairing in the formation of Cooper pairs. Further, 
another promising feature, a Dirac cone like surface state has been observed 
recently in angle resolved photoemission spectroscopy (ARPES) study by 
Hasen {\it et al.}\cite{Hasen}. This non-trivial topological character in the 
superconducting state may pave the path for experimental realization of 
Majorana states in this system. 

Till date there is only one experimental photoemission study reported on 
BiPd\cite{Hasen} which is mainly focused on the surface state bands. In this 
report, our aim is to investigate the near E$_f$ electronic structure of BiPd 
by using ARPES measurements and discuss these results in detail along with our 
calculations based on density functional theory (DFT). We observe a high 
intensity distribution on the Fermi surface (FS) resulting from various 
electron and hole like bands present in the vicinity of E$_f$. These near 
E$_f$ states are composed primarily of Bi-6p with a little admixture of 
Pd-4d$_{x^2-y^2/zy}$ orbitals. The FS mainly consists multi sheets of three 
dimensions (3D) which are due to several spin-orbit splitted bands crossing 
the E$_f$. This 3D character does not favor the nesting between the different 
sheets of FS and weakens the possibility of any density wave instabilities in this system. 
Our results emphasize that the pairing is essentially a spin-single nature 
mediated by the phonons in BiPd.

\section{Experimental and calculation details}

A high quality single crystal of BiPd was synthesized via self flux melt 
growth technique. Stoichiometry was confirmed by using X-ray diffraction 
measurements. Results of the structural and other physical characterizations 
have been reported earlier\cite{awana}. Photoemission spectra were collected 
in angle resolved mode by using a hemispherical SCENTA-R3000 analyzer and a 
monochromatized He source (SCENTA-VUV5040). The photon flux was of the order 
of 10$^{16}$ photons/s/steradian with a beam spot of 2 mm in diameter. Fermi 
energy for all measurements were calibrated by using a freshly evaporated Ag 
film on to the sample holder. The total energy resolution, estimated from the 
width of the Fermi edge, was about 27 meV for the He-I excitation. 
Measurements were performed at a base pressure of $\sim$ 2.0 $\times$ 
10$^{-10}$ mbar at temperature 77 K. A fresh surface of the sample was prepared 
by {\it in-situ} cleaving using post-technique method in the preparation chamber at 
base pressure of 5.0 $\times$ 10$^{-10}$ mbar and the spectra were taken 
within 4.0 hour, so as to avoid any surface degradation. All the measurements 
were repeated many times to ensure the reproducibility of the data. 

First-principles calculations were performed by using a plain wave basis set 
inherent in Quantum Espresso(QE)\cite{qe}. Many electron exchange-correlation 
energy was approximated by Perdew-Burke-Ernzerhof (PBE) 
functional\cite{Ernzerhof,Wang,Chevary}. Both fully relativistic 
ultrasoft\cite{Vanderbilt} and non relativistic norm-conserving 
pseudopotentials were used in the calculations in order to see the SOC 
effects. Fine mesh of k-points with Gaussian smearing of the order 0.0001 Ry 
was used for sampling the Brillouin zone integration, and kinetic energy and 
charge density cut-off were set to 100 Ry and 500 Ry respectively. 
Experimental lattice parameters and atomic coordinates of 
$\alpha$-BiPd\cite{Bhatt}, after relaxed under damped (Beeman) dynamics with 
respect to both ionic coordinates and the lattice vector, were employed in the 
calculations. All parameters were optimized under several convergence tests.

\section{Result and Discussion}

\begin{figure}
\includegraphics[width=8cm,keepaspectratio]{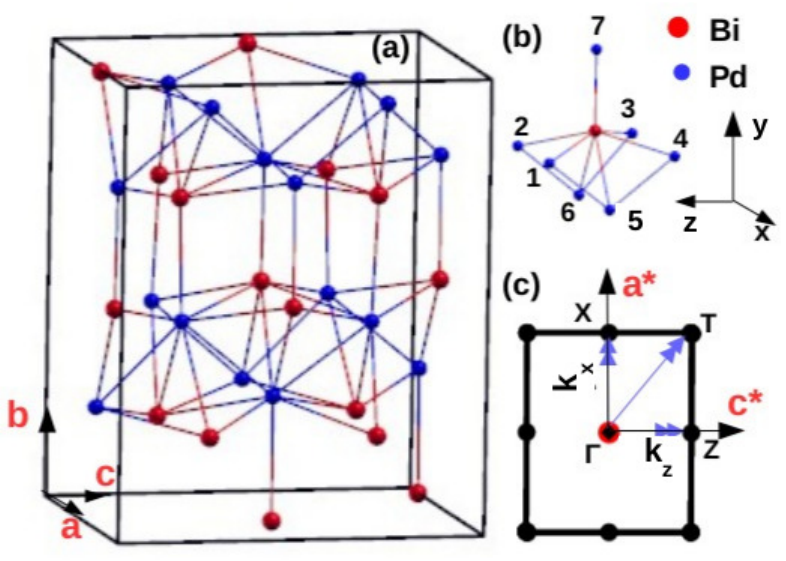}
\caption{\label{crystal}(Color online)(a)  Crystal structure of $\alpha$-BiPd. (b) The coordination environment
of Bi(Pd). (c) The 2D surface Brillouin zone of BiPd where $\Gamma$, Z, X and T are the high symmetry k-points and blue arrows
 denote the k-direction of band mapping.}
\end{figure}

Fig.\ref{crystal}(a) shows the crystal structure of $\alpha$-BiPd, in which 
Bi (Blue) and Pd (Red) atoms are arranged in two adjacent double layers. The 
bonding between alternative layers is weak which makes the (010) plane 
good for cleaving. Bi(Pd) atoms are coordinated to Pd(Bi) atoms situated 
at seven nearest neighbour (nn) sites as shown in Fig.\ref{crystal}(b). The 
average inter-atomic distance of Pd atoms (1-4) from the Bi atom is 2.85 \AA{} 
while the Pd atoms 5 and 6 are at 2.89 \AA{} apart. Similarly, Pd atom 7 
placed on the adjacent double layer is separated by 2.88 \AA{} from the Bi 
atom. Fig.\ref{crystal}(c) present the 2D surface Brillouin zone (BZ), where the high 
symmetry k-points $\Gamma$, X, Z and T are shown along with blue arrows 
indicating the k-directions along which we performed our calculations.

\begin{figure}
\includegraphics[width=9cm,keepaspectratio]{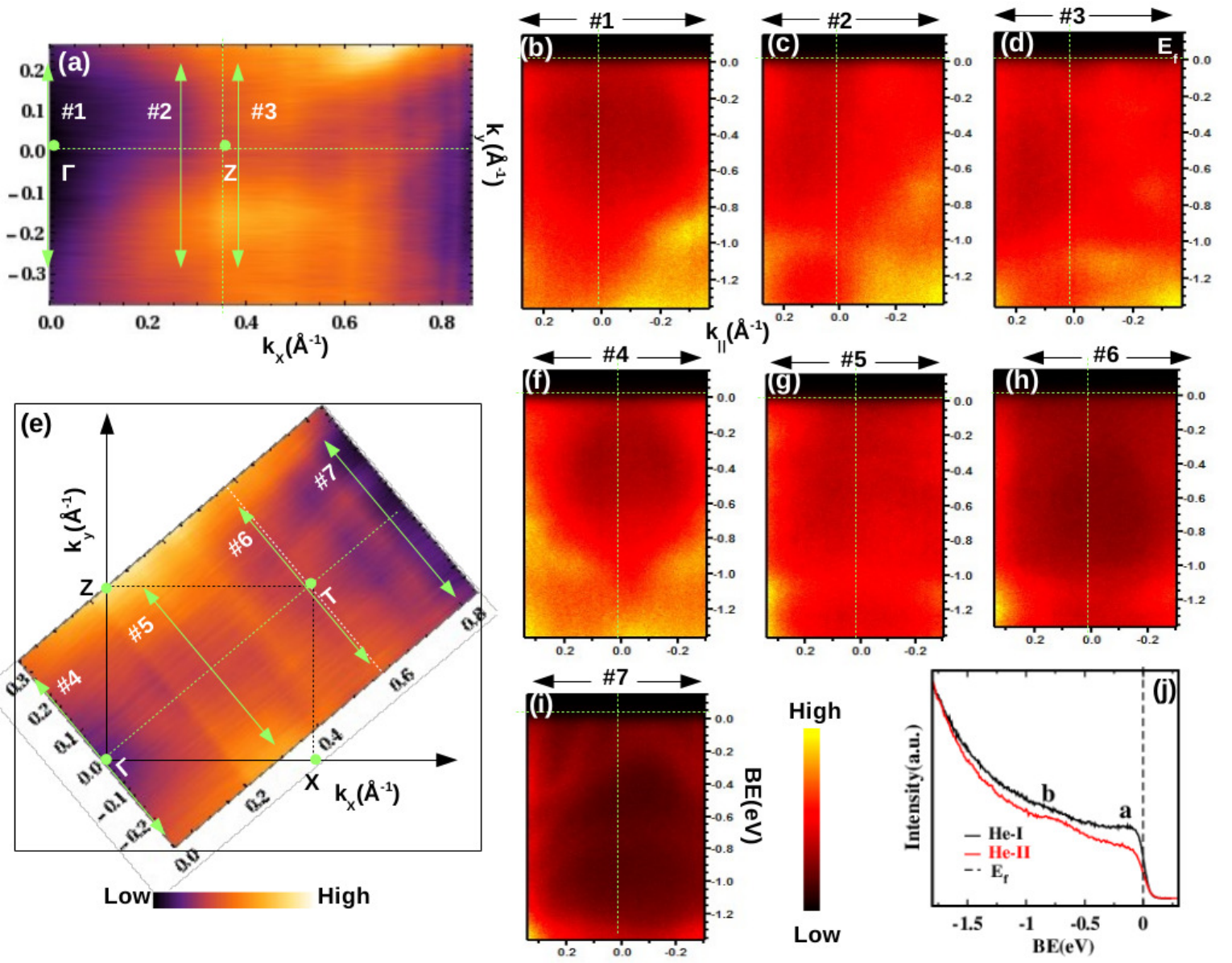}
\caption{\label{expband}(Color online)(a) and (e) ARPES intensity map of FS on $\Gamma$-Z and $\Gamma$-T plane respectively.
(b)-(d )ARPES images of energy dispersion along different cuts \#1 to \#3 and (f)-(i) same images taken along
cuts \#4 to \#7 respectively. (j) Comparison of angle integrated valence band spectra of BiPd
collected at He-I(black) and He-II(red) photon energy.}
\end{figure}

Fig.\ref{expband}(a) shows the ARPES intensity map of the FS of BiPd on the 
$\Gamma$-Z plane. It can be clearly seen from this plot that intensity distribution 
varies significantly at different k-points and it is substantially 
less around the $\Gamma$ point. This distribution gradually increases as we 
move along the $\Gamma$-Z direction and quite high intensity is observed 
around the Z point. The images of energy dispersion along different cuts from 
\#1 to \#3 are displayed in Fig.\ref{expband}(b)-(d) respectively. In 
Fig.\ref{expband}(b), a high intensity region can be found at $\sim$ E$_b$ 
= -0.9 eV and k$_{||}$ = -0.32 \AA{}$^{-1}$ which rapidly disperses towards the 
higher binding energy as k approaches to the $\Gamma$ point. Similarly, moderate 
intensity is observed near the E$_f$ and around E$_b$ = -0.6 eV which also 
shows a strong dispersion. Intensity gets modified remarkably along the other 
two cut directions (\#2 and \#3). Here, intense patches of intensity 
distribution can be found, particularly around the center k point in the 
vicinity of E$_f$ in comparison to cut \#1. Likewise, high intensity regions 
are also observed in many parts on the $\Gamma$-T plane of the FS as shown in 
Fig.\ref{expband}(e). Again, the intensity  distribution is quite 
low around the $\Gamma$ point, whereas it is moderate around T point. Band 
dispersion along cut \#4 to \#7 is displayed in Fig.\ref{expband}(f)-(i) 
respectively. In Fig.\ref{expband}(f), two parabolically dispersing bands can 
be viewed in the higher BE region (E$_b$ = -0.7 to -1.1 eV). Similarly, places 
of high intensity are also observed close to the E$_f$  at $\sim$ k = $\pm$ 0.22 
\AA{}$^{-1}$. These intensity distributions evolve drastically on scanning the 
different cuts along the $\Gamma$-T direction. In the image of cut \#5, 
bands are more intense in the BE range $\sim$ E$_b$ = -0.2 to -0.6 eV 
relative to cut \#4 and again become less intense  in the next 
images collected along the cut \#6 and \#7. The observed high intensity 
distribution at various parts of both the FS plots, which results from the 
different electronic bands,  confirms the good metallic 
character of BiPd. This observation is also  consistent with the resistivity 
behavior of BiPd\cite{Bhanu}. Broadening of the bands which lead to intensity patches 
in the energy dispersion plots could be an outcome of the large number of 
closely spaced bands which are not well resolved in these ARPES images or the 
sizable contribution coming from interlayer 3D coupling\cite{bro}. Nevertheless, 
the strongly dispersive nature of different intensity patterns indicate a 
weakly correlated character of BiPd. In order to elucidate the nature of the 
near E$_f$ states a comparison of angle integrated valence band photoemission 
spectra are presented in Fig.\ref{expband}(j) taken at He-I (21.2 eV) and 
He-II (40.8 eV) photon energy respectively. In He-I spectra (Black), two 
features 'a' and 'b' are visible positioned close to the E$_f$ and at $\sim$ 
E$_b$ = -0.9 eV respectively. This near E$_f$ feature (a) is suppressed in 
He-II spectra (Red) while the intensity of the higher BE feature (b) is 
slightly enhanced in comparison to the He-I case. These changes in the 
spectral weight signify that the near E$_f$ bands (a) could be mainly 
composed of Bi-6p whereas Pd-4d contribution is dominant in the higher BE 
feature (b) (atomic photoinoinization crosssections\cite{lindau} of Bi-6p and 
Pd-4d are different for He-I to He-II).

\begin{figure}
\includegraphics[width=10cm,keepaspectratio]{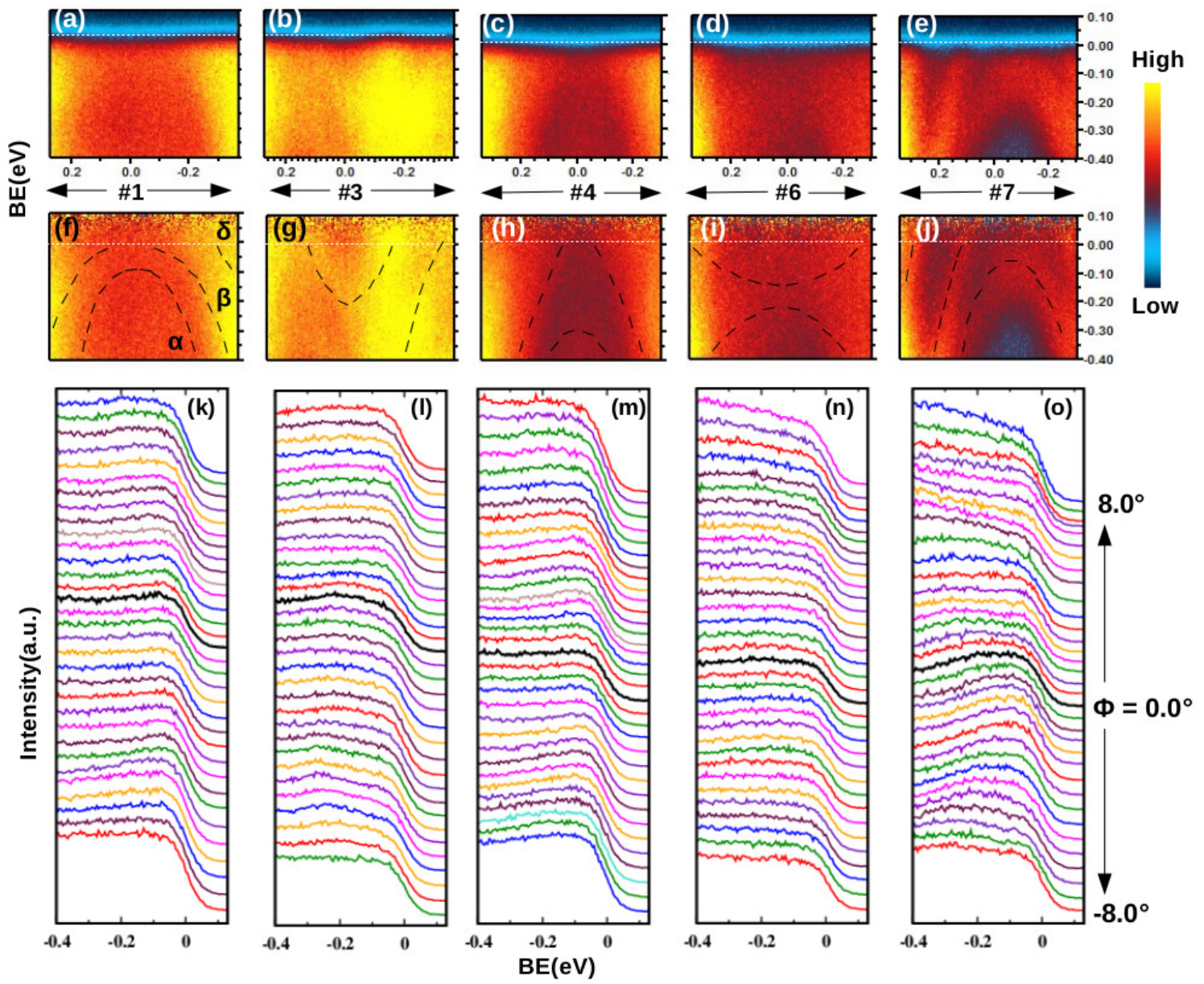}
\caption{\label{edccom}(Color online)(a)-(e) show near E$_f$ region of ARPES intensity plot along 
cut\#1, \#3, \#4, \#6 and \#7 and   corresponding renormalized images by angular integrated 
spectrum  are shown in (f)-(j)  respectively. In (k)-(o) energy density curves(EDC) 
extracted from (a)-(e) are displayed }
\end{figure}

In order to get a deeper insight into the underlying physics behind the near 
E$_f$ bands, an enlarged view of this region in the Fig.\ref{expband}(b) and 
(d) are shown in Fig.\ref{edccom}(a) and (b) respectively. Similarly, 
Fig.\ref{edccom}(c), (d) and (e) correspond to the near E$_f$ zoomed images 
of the Fig.\ref{expband}(f), (h) and (i) respectively. In these plots large 
spectral weight can be seen which indicates the presence of various bands 
crossing the E$_f$. In order to resolve this crossings of near E$_f$ bands  more 
clearly, images Fig.\ref{edccom}(a)-(e) are renormalized by dividing their 
angular integrated spectrum and presented in Fig.\ref{edccom}(f)-(j) 
respectively. In Fig.\ref{edccom}(f) three bands can be identified; two of 
them hole like ($\alpha$ and $\beta$) and one electron like ($\gamma$).
The apex of $\alpha$ band lies $\sim$ E$_b$ = -0.25 eV at the $\Gamma$ point while the band 
$\beta$ crosses the E$_f$ at $\sim$ k$_{||}$ = $\pm$ 0.1 \AA{}$^{-1}$ and 
leads to a hole pocket around the $\Gamma$ point. On the other hand, small 
traces of the electron like band $\gamma$ is found at $\sim$ k$_{||}$ = 0.3 
\AA{}$^{-1}$ which probably forms an electron pocket around the X point. 
Similarly, intensity distribution in the vicinity of the Z point 
(Fig.\ref{edccom}(g)) shows the presence of a one electron like band around 
the Z point and a hole like band touching the E$_f$ nearby the Z point. 
Whereas, along the $\Gamma$-T direction two hole like bands($\alpha$' and $\beta$'), similar to the
bands along the $\Gamma$-X direction are observed around the $\Gamma$ point as
 clear from the Fig.\ref{edccom}(h). The apex of one band($\beta$') is at higher BE (E$_b$ $\sim$ -0.3 eV) 
while the other band($\alpha$'), which shows nearly a linear dispersion carves a 
hole pocket around the $\Gamma$ point. This intensity pattern changes as traversing along this
$\Gamma$-T direction as clear from the Fig.\ref{edccom}(i) and (j) which correspond to ARPES images
taken along the cut \#6 and \#7 respectively. Though, intensity of bands
are not prominent along the cut \#6, three hole like bands could be clearly
seen along the cut \#7. Two of them form hole pockets and the other band jsut 
touches the E$_f$.  The difference in the band dispersion are also evident in energy 
density curves (EDC) plots as shown in Fig.\ref{edccom}(k)-(o) which are 
extracted from the ARPES images Fig.\ref{edccom}(a)-(e) respectively. In these 
EDC plots, sufficient spectral weight is seen close to the E$_f$ and 
its variation with respect to $\phi$ differs along various cuts.

\begin{figure}
\includegraphics[width=9cm,keepaspectratio]{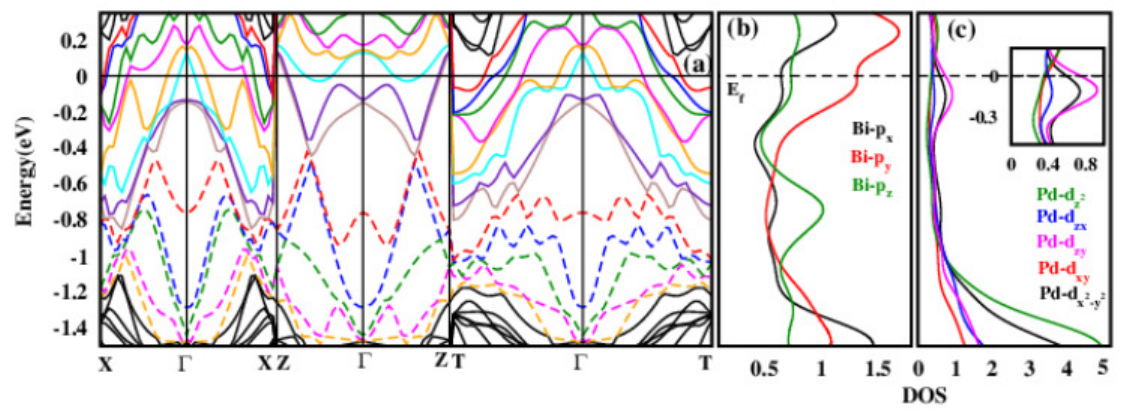}
\caption{\label{calband} (Color online) (a) Non relativistic band structure plot of BiPd.
 (b) and (c)   DOS of individual B-6p and Pd-4d orbitals integrating from all
 the Bi and Pd atoms respectively.}
\end{figure}

We also performed first principles calculations to better understand our 
ARPES findings. In Fig.\ref{calband}(a) non-relativistic band structure of 
BiPd is presented where four electron pockets (red, blue, green and magenta) 
and two hole pockets (orange and cyan) are found around the X and $\Gamma$ 
points respectively. The dispersion of the near E$_f$ bands are slightly 
modified along the $\Gamma$-T direction in comparison to the $\Gamma$-X 
direction, however, the number of electron and hole pockets remain same. On 
the other hand, along the $\Gamma$-Z direction, two almost degenerate (indigo 
and brown) hole like bands cross the E$_f$ close to the Z point leading to 
small hole pockets around the Z point. Moreover, one tiny electron pocket can 
also be observed between the $\Gamma$ and the Z point formed by the cyan band. 
These results are consistent to the resistivity measurements\cite{awana} 
predicting the moderate electronic correlation  in BiPd by estimating 
Kadowaki-Woods value. To figure out the orbital character of these bands, orbital resolved DOS is 
calculated on the same k-space along which bands are obtained and smoothed by 
a Gaussian smearing. These results show that the DOS 
originate from Bi-6p and Pd-4d dominate the near E$_f$ region and DOS of 
different 6p and 4d orbitals of individual Bi and Pd atoms are almost 
identical. So, in Fig.\ref{calband}(b) and (c) DOS of individual Bi-6p and 
Pd-4d orbitals integrating from all the Bi and Pd atoms are displayed. It can 
be clearly seen that Bi-p$_y$ contribution to the near E$_f$ states is quite 
large in comparison to Bi-p$_{x/z}$ as well as sufficient weight of 
Pd-d$_{zy/x^2-y^2}$(see the inset of Fig.\ref{calband}(c)) states are also present 
in this region. Comparing this plot to band picture, it is clear that the bands lying deeper in BE 
(E$_b$ $\sim$ -0.4 to -1.0 eV) red, blue and green (dashed line) and showing 
highly dispersive nature are mainly originated from different Bi-6p orbitals 
while bands E$_b$ $\sim$ -1.2 eV (dashed magenta and orange) are mainly formed 
by Pd-4d$_{x^2-y^2/zy}$ orbitals. These results of band characteristics are 
consistent with the changes seen in the experimental spectral features 'a' and 
'b' as moving from He-I to He-II excitation energy Fig.\ref{expband}(j)). The 
electron pockets (red and blue) around the X point is dominated by Bi-6p$_x$ 
orbital while hole pockets (orange and cyan) around the $\Gamma$ point are 
mainly composed of Bi-6p$_y$ orbital which is different from Bi-6p$_z$ 
character of hole pockets (indigo and brown) around the Z point. Similarly, 
the additional increment of Pd-4d$_{zy}$ and Pd-4d$_{zx}$ orbital character in 
the electron like bands around the T point is the possibly reason for 
modification of these bands in comparison to the same bands seen around the X 
point. These different nature of the near E$_f$ bands, which are predominant 
by different Bi-6p orbitals could be associated to the coordination geometry 
of Bi atom as shown in Fig.\ref{crystal}(b). It is clear that one of the lobe 
of Bi-6p$_y$ orbital is directed towards  4d$_{x^2-y^2}$ orbital of Pd atom 
(7) and the other side has proximity with 4d$_{yx}$ orbital of Pd (5 and 6). 
On the other hand, Pd atoms (1-4) form a geometry like a square planar 
coordination around the Bi atom which is tilted by 30$^\circ$ from the ZX 
plane. These arrangements and nearly equal inter-atomic distances of the Bi 
with these nn Pd atoms favor a strong intermixing between the different Bi-6p 
and Pd-4d orbitals and resulting bands span in the large energy range with 
varying contribution from the different orbitals.  

\begin{figure}
\includegraphics[width=9cm,keepaspectratio]{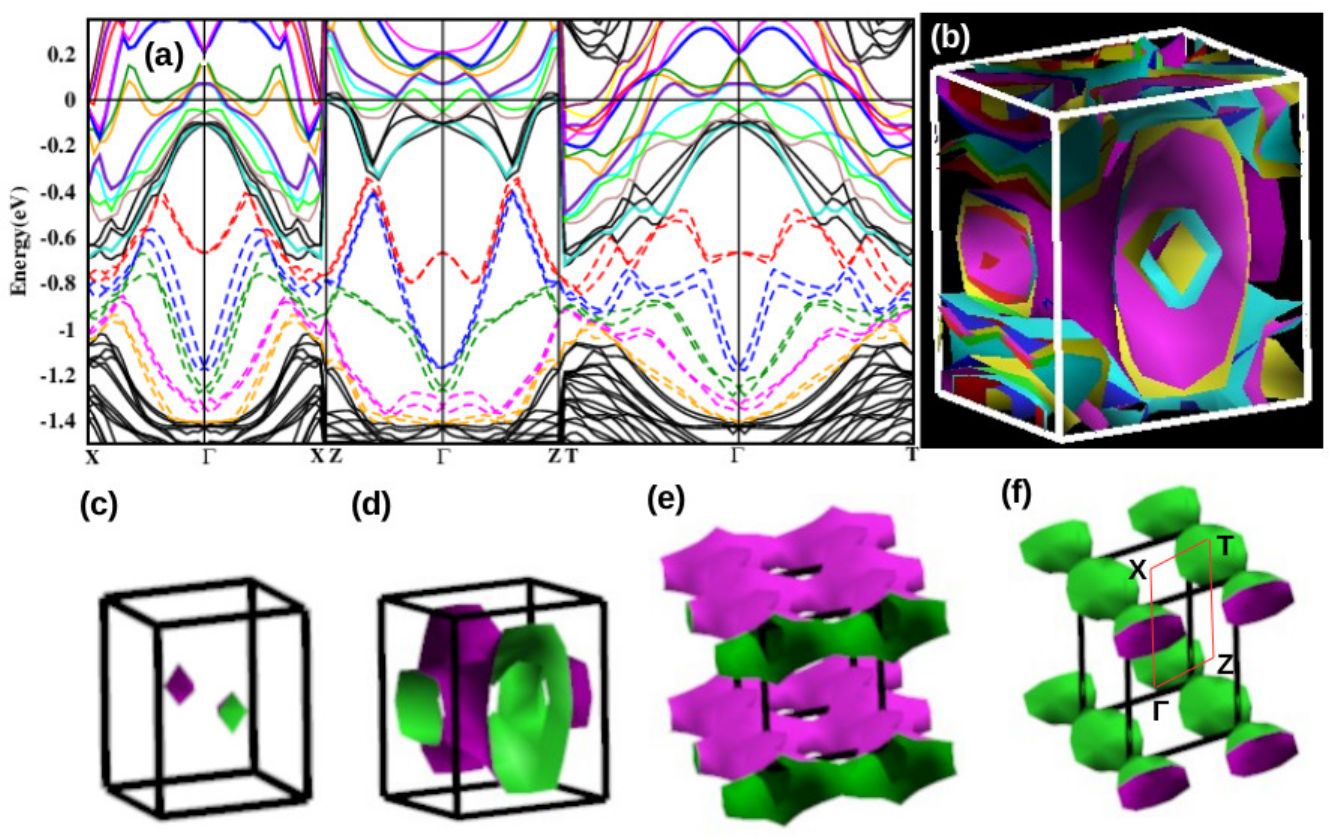}
\caption{\label{calband1} (Color online) (a) Band structure plot of BiPd with including the SOC effects.
(b) composite FS of BiPd. (c)-(f) Individual sheet of FS corresponding to  calculated bands
 turquoise, indigo, blue and violet of (a) respectively.}
\end{figure}

To investigate the effects of SOC, band structure with inclusion of the SOC is 
calculated and illustrated in Fig.\ref{calband1}(a). As clear from this plot 
that the various bands are splitted due to spin orbit interaction and the 
scale of splitting is varied at different k-points. Significant modification 
is observed in the bands which are placed in the vicinity of E$_f$. Several 
new electron and hole pockets with different sizes appear in comparison to the 
non SOC case (Fig.\ref{calband}(a)) Like, one hole pocket (orange band) 
between the $\Gamma$ and X point, two electron pockets (green and brown band) 
between the $\Gamma$ and Z point and electron pockets (maroon and yellow band) 
around the T point are the additional pockets. Moreover, the hole pocket 
shrinks remarkably around the Z point. These changes in the band structure 
signify that the SOC is quite pronounced and play an important role in 
constructing the FS topology in this system. There are many (18) spin-orbit 
split bands cross the E$_f$ which give rise to a complex FS of BiPd as 
depicted in Fig.\ref{calband1}(b). Shape of the FSs originating from different 
spin-orbit split bands is nearly identical. Mainly four types of distinct 
FS sheets can be identified which are shown in Fig.\ref{calband1}(c), (d), 
(e), and (f) corresponding to the calculated band turquoise, indigo, blue and 
violet respectively. Two tiny electron pockets are visible at the center of 
zone faces in the FS made by turquoise band(\ref{calband1}(c)) while two 
connected hole cones, one along k$_z$ (larger in size) and second along k$_y$ 
(smaller in size), are observed in the second FS sheet(\ref{calband1}(d)). On 
the other hand, hole like FS sheet originated from the Bi-6p$_{y}$ (blue) 
dominated band(\ref{calband1}(e)) extends along the k$_y$ and k$_z$ directions 
in a cylindrical tubular form. Similarly, the fourth FS sheet (\ref{calband1} 
(f)) is composed of disconnected pieces of dumble shaped structure which are 
located at the edges of the BZ boundary. This type of FS comprising of several sheets of multidimensional 
character has also been witnessed in similar kind of SC BiPd$_2$\cite{ir} as 
well as some other NCS SCs Ca(Pt/Ir)Si$_3$\cite{vv}, LaPdSi$_3$\cite{mj} and 
Re$_{24}$(Nb;Ti)$_5$\cite{mj1}.

Comparing these results to ARPES findings, we find that the spectral weight distribution
 around the Z(Fig.\ref{calband1}(e)) and T(Fig.\ref{calband1}(f)) points in
comparison to the $\Gamma$ point which is consistent with the experimental FS plots
where high intensity distribution is found around the Z and T points compared to
the $\Gamma$ point(Fig.\ref{expcom}(a) and (e)). Likewise, in band structure along the $\Gamma$-X direction(Fig.\ref{calband1}(a)) 
the blue bands meet the $\Gamma$ point at E$_b$ = -1.17 eV and disperse strongly towards lower BE E$_b$ = -0.6 eV accompanyed
by the green bands. In the ARPES image  Fig.\ref{expcom}(b), intensity pattern shows the identical
behaviour that is higly dispersive in the same BE range and crosses the Gamma point $\sim$
E$_b$ = -1.15 eV.  Further, the calculated red(dashed) band dispersion in the BE between E$_b$ = -0.7 eV to
-0.48 eV is similar to the intensity variation observed in the same BE range in
ARPES data. The calculated bands lying in deeper BE(dashed green, blue, magenta)
along the $\Gamma$-T(Fig.\ref{calband1}(a)) direction show resemblence with the ARPES results(Fig.\ref{expcom}(f)), like
the $\Gamma$-X direction. However, in this image Fig.\ref{expcom}(f) the change in the calculated red(dashed) band  compared
 to the $\Gamma$-X direction is not identified. To consider the near E$_f$ region, along the $\Gamma$-X direction the intensity
pattern marked as an $\beta$ band in (Fig.\ref{edccom}(f)) appears to be the composite result of the
calculated bands(violet, cyan), which form hole pockets around the $\Gamma$ point and
green, brown(Fig.\ref{calband1}(a)). Similarly, the dark green and orange electron
like bands matches to the $\delta$ band(Fig.\ref{edccom}(f)) whereas  the top of closely spaced cyan and black
bands concides with energy postion of $\alpha$ band in ARPES data(Fig.\ref{edcocm}(f)) after renormalized by a factor
of 2.5. This value of renormalization is consistent with the resistivity measurements
predicting the moderate electronic correlation  in BiPd by estimating 
Kadowaki-Woods value\cite{awana}. Same scenario could also be seen along the $\Gamma$-T direction where composite
structure of the calculated  green, brown(Fig.\ref{calband1}(a)) bands possibly leads to the $\beta$' band and renormalized cyan and
black bands resulting the $\alpha$' structre in ARPES image(Fig.\ref{edccom}(h)) taken along the $\Gamma$-T direction. 
Though, the bands are not clearly discernable in the ARPES data but compostie intensity behaviour
show a qualitative agrrement with some of the calculated  bands, particularly in the
higher BE region. These results are helpful for further detailed electronic structre calculations
in order to commensurate the DFT bands to ARPES results. 

As it is clear from our band structure calculations and shown by our ARPES 
measurements (Fig.\ref{edccom}(a)-(e)) various bands containing mainly Bi-6p 
with a little admixture of Pd-4d character are involved in the FS crossings. 
This highlights the possibility of multiband effects in this system, however, 
the strong 3D character of the different FS sheets weakens the possibility of 
nesting between them and consequently existence of any density wave 
instabilities, like charge density wave (CDW) and spin density wave (SDW) are also
less probable in this compound. Thus, the nature of Cooper pairing is expected to be more like 
spin-singlet assisted by phonons which has also been inferred previously from 
different experiments\cite{Bhanu,Sun}. Interestingly, there are a few reports 
which explained these experimetal observations in context of the coexisting 
spin-singlet and triplet pairing due to presence of ASOC in this 
compound\cite{Bhanu1,Mintu,Jiao}. Since, the ASOC split bands occur with 
different spin rotation because of interband SOC, the triplet pairing may be 
favored only at specific places in the resulting FS of such bands, like in 
other NC SC LaPtSi\cite{rogl}. Hence its contribution could be quite less. 
Therefore, BiPd could be essentially a multiband s-wave SC.

\section{Conclusion}

In our ARPES study of BiPd, we found that various bands are involved in the 
crossings of E$_f$ along both the $\Gamma$-X and $\Gamma$-T directions. The FS 
depicts a high intensity distribution at various parts of the surface BZ 
which is consistent with the high metallic nature of BiPd. One hole pocket 
around the $\Gamma$ point and an electron pockets around the X point are 
also identified from the near E$_f$ ARPES intensity plots. These results show 
a fairly good agreement with the calculated band structure, mainly in the higher
BE region, though the bands are nor very discernable as predicted in the calculations.
Our orbital resolved DOS calculation reveals that the near E$_f$ states are primarily composed of Bi-6p orbitals 
with a little admixture of Pd-4d$_{x^2-y^2/zy}$ while the states at higher BE 
($\sim$ E$_b$ = -1.2 eV) are dominated by Pd-4d orbital character. This near 
E$_f$ region is significantly modified with the inclusion of SOC effects and 
various new hole and electron pockets arising from the spin-orbit split 
bands appear in comparison to the non-relativistic case. FS manifested by 
these bands consists of multi sheets of different dimensions, mainly three 
dimensions which  disfavor the nesting conditions and weakens the possibility for any density wave 
instabilities in this system. Since, spin split bands in AOSC driven systems 
have different spin rotation restricting the spin-triplet pairing at specific 
parts of the FS, like in NCS SC LaPtSi, the pairing should mainly be of 
singlet nature mediated via phonons.

\end{document}